\documentclass[epj]{svjour}
\usepackage{graphics}
\usepackage{amsmath,amssymb}
\usepackage{fleqn,epsfig}
\newcommand{\bc}{\begin{center}}
\newcommand{\ec}{\end{center}}
\newcommand{\be}{\begin{equation}}
\newcommand{\ee}{\end{equation}}
\newcommand{\ba}{\begin{array}}
\newcommand{\ea}{\end{array}}
\newcommand{\bea}{\begin{eqnarray}}
\newcommand{\eea}{\end{eqnarray}}
\newcommand{\bt}{\begin{tabular}}
\newcommand{\et}{\end{tabular}}

\newcommand{\bsl}{\boldsymbol}
\newcommand{\ov}{\overline}

\begin{document}
\title{String Correction for Baryon
Orbital Excitations}

\author{O.N.Driga \and I.M.Narodetskii \and A.I.Veselov
\thanks{\emph{This work was supported by RFBR grants
08-02-00657, 08-02-00677, and 09-02-00629.
}
}%
}                     
\offprints{I.M.Narodetskii}          
\institute{Institute of Theoretical and Experimental Physics,
117218 Moscow, Russia}
\date{Received: date / Revised version: date}
%
\abstract{ The correction to the string junction three-quark
potential in a baryon due to the proper moment of inertia of the
QCD string is calculated. The magnitudes of the string corrections
in $P$ wave heavy baryons are estimated.
} 
\maketitle
\section{Introduction}
The formation of an extended object, the QCD string, between the
color constituents inside hadrons plays crucial role in
understanding their properties \cite{DKS:1994}. In a
quark-antiquark system, the string potential $V(r)=\sigma\,r$,
where $\sigma$ is the string tension, represents only the leading
term in the expansion of the QCD string Hamiltonian in terms of
angular velocities. The leading correction in this expansion is
known as a string correction. This correction works well at small
orbital excitations, when there are no long strings. So far, the
string correction was calculeted for the orbitally excited
heavy-light mesons \cite{KNS:2001} and hybrid charmonium states
\cite{KN:2008}. The opposite case of the large angular excitations
is way more advanced.

In a baryon, the leading term of the string potential is
\be\label{eq:string junction}V_Y(\mathbf{r}_1,\,\mathbf{r}_2,\,
\mathbf{r}_3)\,=\,\sigma\,r_{min},\ee where $r_{min}$ is the
minimal string length corresponding to the Y--shaped
configuration. In this paper, we generalize the result of Ref.
\cite{KNS:2001} and derive the string corrections to the string
junction potential (\ref{eq:string junction}). As an application,
we consider the $P$-wave excitations of heavy baryons.

The plan of the paper is as follow. In section
{\ref{section;review}, we briefly review the derivation of the
string potential as applied for low orbital excitations of
quark--antiquark systems. Section 3 is devoted to the
generalization of this derivation for the baryon Wilson loop.
Section 4 presents the numerical results for heavy  baryons.
Finally, Sect.5 summarizes our conclusions. Some details of the
calculation of the baryon string correction are explained in
Appendix.
\section{String $Q\ov Q$ potential}\label{section;review}
First, we recall how the string interaction arizes in a
quark-antiquark system. We create a gauge invariant ~quark-
antiquark state at time $T\,=\,0$ which is annihilated at a later
time T. It is usually assumed that the string potential appears
from the minimal area law asymptotic for an isolated Wilson loop
\bea&& <W>\,=\,\nonumber\\&&=\,\left\langle
TrP\exp{\left(ig{\large\oint_C}dz_{\mu}A_{\mu}\right)}\right\rangle\sim\,
\exp{(-\sigma S_{\rm min})}, \label{eq:W} \eea where the left-hand
side is the average over the stochastic QCD vacuum \cite{D:1987},
and $S_{\rm min}$ is the area of the minimal surface bounded by
the contour $C$ swept by the quark and antiquark trajectories, \be
V_{string}\,=\,-\lim_{T\to\infty}\,\frac{1}{T}\ln<W>\ee

In terms of functions $w_{\mu}(\tau,\beta)$ which map a position
in the parameter space ($\tau,\,\beta$ ) to a point in space-time
the minimal area $S_{\rm min}$ in Eq. (\ref{eq:W}) is written as
\be S_{\rm
min}=\int\,d\tau\int\,d\beta\sqrt{(\dot{w}w')^2-\dot{w}^2w'^2},
\ee where $\dot{w}_{\mu}=\partial w_{\mu}/\partial \tau$,
$w'_{\mu}=\partial w_{\mu}/\partial \beta$. We use  the
instantaneous approximation $t_1=t_2=t$ and the laboratory gauge
$\tau=t$. Choosing for $w_{\mu}(t,\beta)$ the linear form
corresponding to straight-line string \be
w_{\mu}(t,\beta)=\beta\,x_{1\mu}(t)+(1-\beta)\,x_{2\mu}(t),\ee one
obtains for the Lagrangian

\bea
&&L=-\frac{m_1^2}{2\mu_1}-\frac{m_2^2}{2\mu_2}-\frac{\mu_1+\mu_2}{2}+
\frac{\mu_1\dot{\bsl{r}}_1^2}{2}+\frac{\mu_2\dot{\bsl{r}}_2^2}{2}-\nonumber\\[2mm]
&&-\sigma
r\int_0^1d\beta\sqrt{1-[\bsl{n}\times(\beta\dot{\bsl{r}}_1+(1-\beta)\dot{\bsl{r}}_2)]^2},
\label{eq:Lagrangian} \eea where $\bsl{r}=\bsl{r}_1-\bsl{r}_2$,
$\bsl{n}=\bsl{r}/r$, $\beta$ is the remaining internal parameter
along the string, the dots denote derivatives with respect to the
physical time $t$, and $m_i$ are the bare quark masses. The
angular velocities  in (\ref{eq:Lagrangian}) describes the
contribution of the proper inertia of the rotating string. The
Lagrangian (\ref{eq:Lagrangian}) is written in the einbein field
form \cite{BVH:1977}; the einbeins $\mu_1$, $\mu_2$ are introduced
to simplify the treatment of relativistic kinematics. In the
center-of-mass (CM) system \bea &&{\bsl r}_1\,=\,\frac{{\bsl
r}}{2}\,,\,\,\,\,{\bsl r}_2\,=\,-\,\frac{{\bsl
r}}{2},\nonumber\\&& \dot{\bsl r}_1=\frac{\bsl
p_1}{\mu_1},\,\,\,\,\dot{\bsl r}_2=\frac{\bsl p_2}{\mu_1},\,\,\,\,
{\bsl p}_1\,=\,-{\bsl p}_2\,=\,{\bsl p}\eea
The general procedure for the transition from the Lagrangian
(\ref{eq:Lagrangian}) to the quantum Hamiltonian  suggested in
\cite{DKS:1994} is rather complicated due to the presence of
square roots in (\ref{eq:Lagrangian}). However, if one is
interested in the low-lying part of the spectrum, the
potential-type regime can be considered.  The leading term
corresponds to the ${\bsl l}$ independent term, in the expansion
of the string potential in powers of ${\bsl l}^2$, where ${\bsl
l}={\bsl r}\times(-i{\bsl\nabla}_{\bsl r})$ is the orbital
momentum operator. The first correction, of order ${\bsl l}^2$, is
known as the string correction.

The Hamiltonian function corresponding to (\ref{eq:Lagrangian})
reads \be\label{eq:Hamiltonian}
H=\sum_{i=1}^2\left(\frac{\bsl{p}^2+m_i^2}{2\mu_i}+\frac{\mu_i}{2}\right)
+V_{string}(\bsl r), \label{eq:H} \ee where \bea &&
V_{string}(\bsl{r})=\nonumber\\&&\sigma
r\int_0^1d\beta\sqrt{1-\left[\bsl{n}\times(\beta\,\frac{\bsl{p}}{\mu_1}-(1-\beta)\frac{\bsl{p}}{\mu_2}\right]^2}
\eea Expanding the square root and making the substitution ${\bsl
p}\to-i{\bsl\nabla}_{\bsl r}$ one obtains an approximate
expression for the string potential
\cite{KNS:2001}\bea\label{eq:string correction}&&V_{\rm
string}(\bsl{r})=\nonumber\\&&=\sigma
r\int_0^1d\beta\left(1-\frac{1}{2}\,\left[\bsl{n}\times(\beta\frac{\bsl
p}{\mu_1}-(1-\beta)\frac{\bsl
p}{\mu_2}\right]^2\right)\nonumber\\&&\approx\sigma r-\frac{{\bsl
L}^2}{2\,r}\int\limits_0^1\left(\frac{\beta}{\mu_1}-\frac{1-\beta}{\mu_2}\right)^2d\beta
 \nonumber\\&&=
\sigma\,r\,-\,\frac{\sigma
(\mu_1^2+\mu_2^2-\mu_1\mu_2)}{6\mu_1^2\mu_2^2}\frac{\bsl{l}^2}{r}
 \label{eq:string}\eea  The last term in Eq.
 (\ref{eq:string correction}) is
the string correction. Its sign is negative so that the
contribution of the string corrections lowers the energy of the
system, thus giving a negative contribution to the masses of
orbitally excited states, leaving those with $l=0$ intact.
\section{The string baryon potential}
 Now let us apply a
similar  consideration to the baryon states.
In tricolor QCD, three quarks are produced at the origin and then
fly, eventually creating strings of the chromo electric field. If
the three quarks are at positions ${\bsl r}_1$, ${\bsl r}_2$, and
${\bsl r}_3$, the strings meet at an interior point, known as the
string junction point, where the angles between the flux tubes are
120$^0$, provided that none of the angles of the triangle formed
by the three quarks exceeds 120$^{0}$. Otherwise, the
configuration of minimal total length is made of two flux tubes
meeting at the third quark location. In the general case time
evolution produces a three-bladed Wilson loop shown in Fig.
\ref{fig:BWL}.

For a baryon with the genuine string junction point the
calculation of the string correction is a very cumbersome problem.
The calculations are greatly simplified, however, if the string
junction point is chosen to coincide with the CM coordinate ${\bsl
R}_{\text cm}$. In this case, the complicated string junction
potential is approximated by a sum of the one--body confining
potentials. The accuracy of this
 approximation for the unperturbed three--quark Hamiltonian was tested in \cite{NSV:2008}. In particular,
 for the $P$--wave baryon states the accuracy
 is better than 1$\%$ .

\begin{figure}
\begin{center}
\epsfxsize=7.5cm \epsfysize=7.5cm \epsfbox{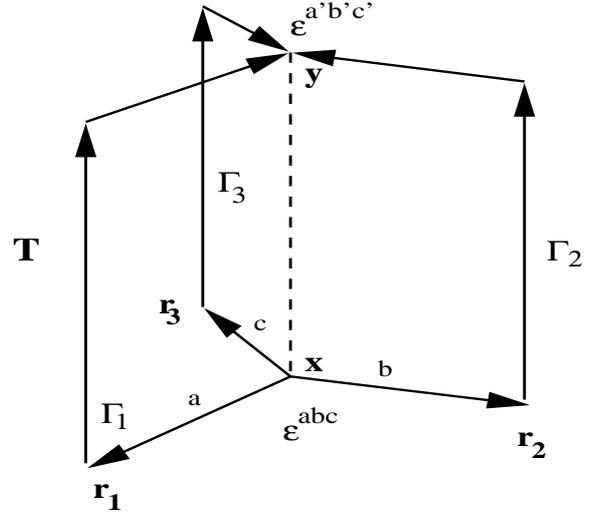}
\end{center}
\caption{The baryonic Wilson loop. The quarks are located at
positions ${\bsl r}_1,{\bsl r}_2$ and ${\bsl r}_3$.}
\label{fig:BWL}
\end{figure}

 Letting ${\bsl R}_{cm}\,=\,0$, we arrive at the following
 generalization of Eq. (\ref{eq:string}) for a baryon
\bea\label{eq:V_string} &&V_{\rm string}^{\rm CM}\,=\,\sigma
\,\sum_i\,|\bsl{r}_i|\,\int^1_0\,
d\beta_i\,\sqrt{1-\frac{\beta_i^2}
 {\mu_i^2\,r_i^2}
 \,{\bsl
 l}_i^2}\,\nonumber\\&&\approx\,\sigma\,\sum\limits_i\,r_i\,\left(1-\frac{1}{6}\,\frac{{\bsl
 l}_i^2}{\mu_i^2\,r_i^2}\right)
, \eea  where~ ${\bsl l}_i\,=\,{\bsl r}_i\,\times\,{\bsl p}_i$ are
the orbital momentum operators of the individual quarks. Each term
in Eq. (\ref{eq:V_string}) is associated with the string beginning
at the center-of-mass coordinate and ending at the quark position
${\bsl r}_i$. As a result, for the string correction we get
\be\label{eq:Delta_M}\Delta M_{\rm
string}\,=\,-\,\frac{\sigma}{6}\,<\Psi|\sum_i\,\frac{{\bsl
l}_i^2}{\mu_i^2\,r_i}\,|\Psi>,\ee where $\Psi$ is an eigenfunction
of the unperturbed Hamiltonian.
\section{The string correction}
The calculation of the matrix element (\ref{eq:Delta_M}) is more
conveniently performed in terms of the Jacoby variables
${\bsl\rho}$, ${\bsl\lambda}$ rather than quark coordinates ${\bsl
r}_i$. The baryon wave function depends on the three-body Jacobi
coordinates \bea\label{eq:Jacobi}
&&\bsl{\rho}_{ij}=\sqrt{\frac{\mu_{ij}}{\mu_0}}\,(\bsl{r}_i-\bsl{r}_j),\nonumber\\&&
\bsl{\lambda}_{ij}=\sqrt{\frac{\mu_{ij,\,k}}{\mu_0}}
\left(\frac{\mu_i\bsl{r}_i+\mu_j\bsl{r}_j}{\mu_i+\mu_j}-\bsl{r}_k\right),\eea
where $\mu_{ij}$ and $\mu_{ij,k}$ are the appropriate reduced
masses:
\begin{equation}\mu_{ij}=\frac{\mu_i\mu_j}{\mu_i\,+\,\mu_j}\,,\,\,\,\,\,\,
\mu_{ij,\,k}=\frac{(\mu_i\,+\,\mu_j)\mu_k}{M}\,,\end{equation}
$M\,=\,\mu_1+\mu_2+\mu_3$, and $\mu_0$ is an arbitrary parameter
with the dimension of mass, which drops out in the final
expressions. The hyperradius
$R^2=\mu_0\,({\bsl\rho}_{ij}^2+{\bsl\lambda}_{ij}^2)$
 does not depend on $\mu_0$ and is
invariant under permutations. There are three equivalent ways of
introducing the  coordinates (\ref{eq:Jacobi}), which are related
to each other by linear transformations with the Jacobian equal to
unity.

The quark orbital momenta ${\bsl l}_i$ in Eq. (\ref{eq:V_string})
can be simply constructed from the orbital momenta corresponding
to different Jacoby variables. It is convenient to choose for each
term containing ${\bsl l}_i^2$ in (\ref{eq:V_string}) an
appropriate set of Jacoby coordinates ${\bsl\rho}_{jk}$, then we
obtain \be \frac{[\,{\bsl r}_i\times{\bsl
p}_i\,]^2}{\mu_i^2\,r_i}\,=\,\left(
\frac{\mu_j+\mu_k}{M}\right)^{3/2}\sqrt{\frac{\mu_k}{\mu_0}}\,\,\frac{1}{\mu_i^2\,\lambda_{jk}}\,{\bsl
L}_{\lambda_{jk}}^2,\label{eq:L}\ee where
$ijk\,=\,123,\,231,\,312,$ and \be {\bsl
L}_{\lambda_{jk}}\,=\,(-i)\,
\left({\bsl\lambda}_{jk}\times\frac{\partial}{\partial\,{\bsl\lambda}_{jk}}\right)\ee


 In what
follows we consider the $nnQ$ baryons, where $n$ stands for the
light quarks $u$ or $d$, and $Q$ stands for the heavy quarks $c$
or $b$. We use the basis in which the heavy $Q$-quark is singled
out as quark $3$ but in which the $nn$ quarks are still
antisymmetrized. The similar basis is used for the $nsQ$ baryons.
These basis states diagonalize the confinement problem with
eigenfunctions that correspond to separate excitations of the
light and heavy quarks. We refer to these excitations as the
$\rho$\,- and $\,\lambda$-excitations, respectively. The
nonsymmetrized $nnQ$ and $nsQ$ bases usually provide a much
simplified picture of the states. The physical P-wave states are
neither pure SU(3) states nor pure $\rho$ or $\lambda$ excitations
but linear combinations of all states with a given $J$. Most
physical states are, however, closer to pure $\rho$ or $\lambda$
states than to pure SU(3) states \cite{IK:1978}.

For definiteness we  now discuss the simplest case of $P$-wave
states. The angular functions of these states can be of two types.
In Cartesian coordinates they are given by \bea&& {\bsl
Y}_{\bsl\rho}=\sqrt{\frac{6}{\pi^3}}\frac{{\bsl\rho}_{12}}{R}=\sqrt{\frac{6}{\pi^3}}\,\left(\alpha_{31}\frac{
{\bsl\rho}_{31}}{R}+\beta_{31}\frac{{\bsl\lambda}_{31}}{R}\right)\nonumber\\&&=\sqrt{\frac{6}{\pi^3}}\left(\alpha_{23}\frac{{\bsl\rho}_{23}}{R}
+\beta_{23}\frac{{\bsl\lambda}_{23}}{R}\right),\label{eq:31}\eea
and\bea&& {\bsl
Y}_{\bsl\lambda}=\sqrt{\frac{6}{\pi^3}}\,\frac{{\bsl\lambda}_{12}}{R}=\sqrt{\frac{6}{\pi^3}}\left(-\beta_{31}\frac{
{\bsl\rho}_{31}}{R}
+\,\alpha_{31}\frac{{\bsl\lambda}_{31}}{R}\right)\nonumber\\&&=\sqrt{\frac{6}{\pi^3}}\left(-\beta_{23}\frac{{\bsl\rho}_{23}}{R}
+\alpha_{23}\,\frac{{\bsl\lambda}_{23}}{R}\right),\eea where the
mass dependent coefficients $\alpha_{ij}$ and $\beta_{ij}$ are
defined in (\ref{eq:31}), (\ref{eq:23}). The functions ${\bsl
Y}_{{\bsl\rho},\,{\bsl\lambda}}$ are the eigenvalues of the
six-dimensional angular operator \be {\cal\bsl L}^2{\bf
Y}_{\bsl\rho,\,{\bsl\lambda}}(\theta,{\bf n}_{\rho},{\bf
n}_{\lambda})\,=\,-\,K(K+4){\bf Y}_{\bsl\rho,\,{\bsl\lambda}}
(\theta,{\bf n}_{\rho},{\bf n}_{\lambda}),\ee where\be{\cal\bsl
L}^2\,=\,\frac{\partial^2}{\partial
\theta^2}\,+\,4\cot\theta\,\frac{\partial}{\partial
\theta}-\frac{{\bf l}_{\rho}^2}{\sin^2\theta}\,-\,\frac{{\bf
l}_{\lambda}^2}{\cos^2\theta}\,,\ee  and $K=1$ is the grand
orbital momentum. The functions $\Psi_{\nu}$ in (\ref{eq:Delta_M})
are written as
\be\Psi_{\nu}({\bsl\rho})\,=\,\frac{u_{\nu}(R)}{R^{5/2}}\,{\bsl
Y}_{\nu}(\theta,{\bf n}_{\rho},{\bf
n}_{\lambda}),\,\,\,\,\,\,\nu={\bsl\rho},\,{\bsl\lambda},\ee where
we have introduced the reduced radial functions $u_{\nu}(R)$.
The further
calculations are elementary and lead to the final
result\footnote{Some details of the calculations are presented in
Appendix}
\bea\label{eq:rho}&&\Delta
M_{\rho}\,=\,-\,\frac{64\,\sigma}{45\pi}\,\frac{1}{\sqrt{M}\,(\mu_1+\mu_2)}\nonumber\\&&\times\left(
\frac{\sqrt{\mu_2+\mu_3}}{\mu_1^{3/2}}\mu_2\,+\,\frac{\sqrt{\mu_1+\mu_3}}{\mu_2^{3/2}}\mu_1
\right)\gamma_{\rho}
 \,,\eea
 \bea\label{eq:lambda}&&\Delta
 M_{\lambda}\,=\,-\frac{64\sigma}{45\pi}\,\frac{\mu_3}{M^{3/2}(\mu_1+\mu_2)}
 \nonumber\\&&\times\left(
\left[\frac{\mu_1+\mu_2}{\mu_3}\right]^{\frac{5}{2}}+\sqrt{\frac{\mu_2+\mu_3}{\mu_1}}+
\sqrt{\frac{\mu_1+\mu_3}{\mu_2}}\right)\gamma_{\lambda},\eea where
\be\gamma_{\nu}=\int\limits_0^{\infty}\frac{\hat
u_{\nu}(x)}{x}dx,\,\,\,\,\,\nu=\rho,\lambda,\ee
$x=\sqrt{\mu_0}\,R,$ and the functions ${\hat u}(x)$ are
normalized as \be \int_0^{\infty}{\hat u}^2(x)dx=1\ee
 Eqs. (\ref{eq:rho}), (\ref{eq:lambda}) define the magnitude of
 the string correction for the $P$-wave baryon states.
The generalization to the $D$-wave excitations is straightforward.

 \section{Numerical results}

Now we are in position to estimate the magnitudes of string
corrections.
Here we concentrate on heavy baryons.

Our calculations of the unperturbed baryon masses and wave
functions are performed using the effective Hamiltonian derived in
the Field Correlator Method \cite{DDSS:2002}. For the different
technical aspects of these calculation see refs.
\cite{NT:2004}--\cite{KNV:2009}. As an unperturbed Hamiltonian we
consider  the Hamiltonian specified in Ref. \cite{NSV:2008}
 \bea
 \label{eq:zero_order_H}
&&H_0=\sum\limits_{i=1}^3\left(\frac
{m_{i}^2}{2\mu_i}+\frac{\mu_i}{2}\right)+\nonumber\\&&\sum\limits_{i=1}^3\frac{{\bsl
p}_i^2}{2\mu_i}+V_Y(\mathbf{r}_1,\,\mathbf{r}_2,\,
\mathbf{r}_3)+V_{C},\eea where the long-range confining force
(\ref{eq:string junction}) is augmented by the short-range Coulomb
potential, \be V_{C}=-\frac{2}{3}\alpha_s\sum_{i\leq
j}\frac{1}{r_{ij}}\ee As in Eq.(\ref{eq:Lagrangian}), an
approximate einbein field method is used for derivation of Eq.
(\ref{eq:zero_order_H}): einbeins $\mu_i$ are treated as c-number
variational parameters. The eigenvalues $E_0(m_i,\mu_i)$ of the
Hamiltonian (\ref{eq:zero_order_H}) are found as functions of the
bare quark masses $m_i$ and einbeins $\mu_i$, and are finally
minimized with respect to the $\mu_i$. With such simplifying
assumptions the spinless Hamiltonian $H_0$ takes an apparently
nonrelativistic form, with einbein fields playing the role of the
constituent masses of the quarks.

The zero-order baryon mass $M_0$ is
\begin{equation}
\label{eq:zero order mass} M_0\,=\,\sum\limits_{i=1}^3\left(\frac
{m_{i}^2}{2\mu_i\,}+
\,\frac{\mu_i}{2}\right)\,+E_0(\mu_i)\,+\,C\end{equation} In Eq.
(\ref{eq:zero order mass}) a constant $C$ is the sum of the
nonperturbative self-energies of the quarks propagating through
the vacuum background field which
%
provides an overall negative shift of $M_0$ required by
phenomenology \cite{S:2001}. Finally, $E_0(\mu_i)$ is an
eigenvalue of the Schr\"{o}dinger equation
\bea\label{eq:se}&&\frac{d^2
u_{\nu}(R)}{dR^2}\,+\nonumber\\&&2\left(E_0-\frac{35}{8R^2}-
V_{\rm Y}^{\nu}(R)-\,V_{\rm C}^{\nu}(R)\right)u_{\nu}(R)=0,\eea
where $V_{\rm Y}^{\gamma}(R)$, $V_{\rm C}^{\gamma}(R)$ are the
averages of the string junction and Coulomb potentials,
respectively, over the six-dimensional sphere $\Omega_6$. Explicit
expressions of these quantities can be found in Ref. \cite{DNV}.
We use the same parameters, $\sigma=0.15$ GeV, $\alpha_s=0.39$,
$m_n=7$ MeV, $m_s=185$ MeV, $m_c=1.36$ GeV, and $m_b=4.71$ GeV,
that were used in Ref. \cite{NTV:2009} for the analysis of the
ground states of heavy baryons. Thus the model is totally fixed.

The wave function factors $\gamma_{\nu}$ in Eqs. (\ref{eq:rho}),
(\ref{eq:lambda}) which define the magnitude of the string
correction do not essentially depend on the baryon flavor nor on
the type of excitation. This property mentioned for the light
baryons in Ref. \cite{KNV:2009} is also valid for heavy baryons.
{\it E.g.} for all considered baryons we get  $\gamma_{\rho},
\,\gamma_{\lambda}\,\sim$ 0.30-0.31 GeV$^{1/2}$, which  are
practically the same values as were found for the light baryons.
The relative size of the string corrections for different baryons
is mainly defined by the constituent quark masses $\mu_i$ that are
themselves the calculable quantities.

Note that our technique does not take into account chiral degrees
of freedom, which are essential for light diquarks in the
$\Lambda_Q$ baryons, so we consider the $\Sigma_Q$ and $\Xi_Q$
where the chiral effects are expected to play the minor role. The
result of the calculation of the $P$-wave states is given in Table
\ref{tab:L=1}. In this table we present the dynamical quark masses
$\mu_n$, $\mu_s$ and $\mu_Q$ (columns 3--5) for various heavy
baryons. The latter are computed solely in terms of the bare quark
masses, $\sigma$ and $\alpha_s$ and marginally depend on a baryon.
We also display the eigenvalues $E_0$ of the Schr\"{o}dinger
equation (column 6), the zero-order masses  calculated without the
string corrections from Eq. (\ref{eq:zero order mass}) (column 7),
the string correction itself (column 8) and the total baryon
masses (column 9). Note that the $\lambda$ excitations are
typically $\sim 100$ MeV heavier than the $\rho$ excitations.

Because in this paper we neglect the spin interactions, it would
be premature to compare literally the results of our piloting
calculations with experimental data for the $\Sigma_Q$ and $\Xi_Q$
excitations. Presently, the Particle Data Group \cite{PDG} lists
only two $\frac{1}{2}^-$ $c$ baryons, $\Sigma_c(2800)$ and
$\Xi_c((2790)$. Spin and parity of these states have not been
measured but were assigned on the basis of quark model
predictions. The $P$-wave $b$ baryons have not been detected yet.
In Table 2 we compare our predictions for the $\bsl\lambda$
excitations with the results of several quark models
\cite{EFG:2007}\,-\,\cite{G:2007}. It is seen that our results
match the region of the predictions well.

\section{Conclusions}
 In this paper, we have shown how the dynamics of the  strings
with quarks at the ends influences the masses of the baryon
orbital excited states.  Note that the string correction is
totally missing in quark models based on the relativistic
equations with local potentials. We extend our approach to baryons
containing charm and bottom quarks and calculate the string
correction to the $P$-wave states of the $nnQ$ and $nsQ$ baryons.
Our main result is given by Eqs. (\ref{eq:rho}),
(\ref{eq:lambda}). The numerical results are given in Table
\ref{tab:masses}. For each baryon, we have calculated the
dynamical quark masses $\mu_i$, the baryon masses with the
self-energy corrections in Eq. (\ref{eq:zero order mass})
and the string corrections. The latter provide an extra piece of
the effective interquark potential, which is entirely due to the
string-type interaction in QCD. We emphasize that no fitting
parameters were used in baryon spectrum calculations.
The estimated string corrections give a negative contribution to
the baryon masses of the order 30-50 MeV. Our comparative study
provides deeper insight into the quark model results for which the
string interactions encode the underlying QCD dynamics.


\section*{Appendix}
\setcounter{equation}{0}
\def\theequation{A.\arabic{equation}}
In this appendix, we present for an illustrative purpose the
explicit calculation of the first term with $i=1$ in Eq.
(\ref{eq:V_string}). Consider as an example the $\bsl\rho$
excitation. Eq. (\ref{eq:L}) implies that
\bea\label{eq:me}&&\langle\Psi_{\rho}|\frac{{\bsl
l}_1^2}{\mu_1^2r_1}|\Psi_{\rho}\rangle=\,\nonumber\\&&\left(
\frac{\mu_2+\mu_3}{M}\right)^{3/2}\sqrt{\frac{\mu_3}{\mu_0}}\,\frac{1}{\mu_1^2}\,\langle\Psi_{\rho}|\frac{1}{\lambda_{23}}{\bsl
L}_{\lambda_{23}}^2|\Psi_{\rho}\rangle\eea  At this point it is
convenient to restore the quark numeration in the Jacobi
coordinates. We use the transformation law  \be{
\left(\begin{array}{r} \bsl\rho_{12} \\[1mm] \bsl\lambda_{12}\\[1mm]
\end{array}
\right)\,=\, \left(
\begin{array}{rr}\alpha_{31}
& \beta_{31}\\[1mm] -\,\beta_{31}
&\alpha_{31}
\end{array} \right)\left(\begin{array}{r}  \bsl\rho_{31}
\\[1mm]
\bsl\lambda_{31}\end{array}\right)},\label{eq:transformation}\ee
\be{
\left(\begin{array}{r} \bsl\rho_{12} \\[1mm] \bsl\lambda_{12}\\[1mm]
\end{array}
\right)\,=\, \left(
\begin{array}{rr}\alpha_{23}
& \beta_{23}\\[1mm] -\,\beta_{23}
&\alpha_{23}
\end{array} \right)\left(\begin{array}{r}  \bsl\rho_{23}
\\[1mm]
\bsl\lambda_{23}\end{array}\right)},\ee with
\bea\label{eq:31}&&\alpha_{31}\,=\,-\,\sqrt{\frac{\mu_2\mu_3}{(\mu_1+\mu_2)(\mu_1+\mu_3)}}\,,\nonumber\\&&\beta_{31}
\,=\,\sqrt{\frac{\mu_1\,M}{(\mu_1+\mu_2)(\mu_1+\mu_3)}}\,,\eea
\bea&&\alpha_{23}\,=\,-\,\sqrt{\frac{\mu_1\mu_3}{(\mu_1+\mu_2)(\mu_2+\mu_3)}}\,,\nonumber\\&&\beta_{23}
\,=\,-\,\sqrt{\frac{m_2\,M}{(\mu_1+\mu_2)(\mu_2+\mu_3)}}\,.\label{eq:23}\eea
Substituting
$\Psi_{\rho}(12;3)=\alpha_{23}\Psi_{\rho}(23,1)+\beta_{23}\Psi_{\rho}(23,1)$
and noting that \be{\bsl
 L}_{\lambda_{23}}^2\frac{{\bsl\rho}_{23}}{R}=0,\,\,\,\,\,\,\,\,\,\,{\bsl
 L}_{\lambda_{23}}^2\frac{{\bsl\lambda}_{23}}{R}=2\,\frac{{\bsl\lambda}_{23}}{R}\ee
 we obtain for the matrix element in the r.h.s. of (\ref{eq:me})
\bea&& \langle\Psi_{\rho}|\frac{1}{\lambda_{23}}{\bsl
L}_{\lambda_{23}}^2|\Psi_{\rho}\rangle\,=\nonumber\\&&=2\beta_{23}\int
\frac
{(\alpha_{23}{\bsl\rho_{23}}+\beta_{23}{\bsl\lambda_{23}}){\bsl
\lambda_{23}}}
{\lambda_{23}}d\Omega\times\nonumber\\&&\int\limits_0^{\infty}\frac{u_{\rho}^2(R)}{R^2}dR=\frac{64\pi^2R}{45}\beta_{23}^2\int\limits_0^{\infty}
 \,\frac{{u}_{\rho}^2(R)}{R}\,dR,\eea
 where we have used
 \bea
 &&\int\frac{\rho_{23}^i\lambda_{23}^i}{\lambda_{23}}d\Omega_6=0,\nonumber\\&&\int\frac{(\lambda_{23}^i)^2}{\lambda_{23}}d\Omega_6\,=\,
 \frac{1}{3}\int\lambda_{23}d\Omega_6\,=\,\frac{32\pi^2R}{45}.\eea
As the result, we obtain
 \bea
&&\langle\Psi_{\rho}|\frac{{\bsl
l}_1^2}{\mu_1^2r_1}|\Psi_{\rho}\rangle=\,\nonumber\\&&\frac{384}{45\pi}\,\frac{\mu_2}{\sqrt{M}\,\sqrt{\mu_2\,+\mu_3}\,\mu_1^{3/2}}
\int\limits_0^{\infty}
 \,\frac{{u}_{\rho}^2(R)}{R}\,dR,\eea
which corresponds to the first term in Eq. (\ref{eq:rho}). An
analogous calculation with the obvious changes leads to the second
term in (\ref{eq:rho}). The string correction to the
${\bsl\lambda}$ excitation can be evaluated in a similar way.

\newpage

\begin{table*}[t]
 \caption{Heavy baryons with $L\,=\,1$}
 \label{tab:L=1} \vspace{2mm}

\centering
\begin{tabular}{ccccccccc} \hline\hline\\
Baryon&~~${\bf L}_{\alpha}$ &~~$\mu_n$&~~ $\mu_s$&~~$\mu_h$&~~
$E_{0}$&$M_0$&$\Delta M_{\rm string}$& $M_B$\\ \\ \hline\hline\\
$nnc$&${\bf 1}_{\rho}$&536&&1452&1397&2920&-48&2872\\  $nnc$&${\bf
1}_{\lambda}$&495&&1491&1377&2832&-36&2796
\\
$nsc$& ${\bf 1}_{\rho}$&542&582&1455&1372&2954&-45&2909\\
$nsc$& ${\bf 1}_{\lambda}$&497&544&1494&1353&2867&-33&2834\\
\\

\hline\\
$nnb$&${\bf 1}_{\rho}$&570&&4746&1294&6240&-46&6194\\  $nnb$&${\bf
1}_{\lambda}$&540&&4764&1234&6132&-41&6091
\\
$nsb$&${\bf 1}_{\rho}$&574&615&4748&1271&6272&-43&6229
\\
$nsb$&${\bf 1}_{\lambda}$&542&588&4765&1211&6164&-38&6126
\\ \\
\hline\hline \label{tab:masses}
\end{tabular}
\vspace{2mm}
\end{table*}
\begin{table*}[t]
 \caption{Theoretical predictions for masses of heavy baryons
 in various quark models} \label{comparison} \vspace{2mm}

\centering
\begin{tabular}{cccccccc} \hline\hline\\
&This work&$J^P$&\cite{EFG:2007}
&\cite{CI:1986}&\cite{M:2006}&\cite{G:2007}&PDG \cite{PDG}
\\\\
\hline\hline\\
$\Sigma_c$&2800&$\frac{1}{2}^-$&2795&2765&2769&2706&2802$\,(?)$\\
&&$\frac{1}{2}^-$&2805&2770&2817&2791&2766$\,(?)$\\
&&$\frac{3}{2}^-$&2761&2770&2799&2706&\\
&&$\frac{3}{2}^-$&2799&2805&2815&2791&\\\\
$\Xi_c$&2834&$\frac{1}{2}^-$&2801&&2769&2749&2790\\
&&$\frac{1}{2}^-$&2928&&&2829&\\
&&$\frac{3}{2}^-$&2820&&2771&2749&\\
&&$\frac{3}{2}^-$&2900&&&2829&\\\\
$\Sigma_b$&6091&$\frac{1}{2}^-$&6108&6070&&6039\\
&&$\frac{3}{2}^-$&6076&6070&&&\\\\
$\Xi_b$&6125&$\frac{1}{2}^-$&6110&&&6076&\\
&&$\frac{3}{2}^-$&6130&&&6076&\\\\

\hline\hline
\end{tabular}
\vspace{2mm}

\end{table*}
\end{document}